\documentclass[12pt]{article}
\usepackage{amsmath}
\usepackage{bm}
\usepackage{color}

\begin{document}
\begin{center}
\begin{large}
{\bf Rotationally invariant noncommutative phase space of canonical type with recovered weak equivalence principle}
\end{large}
\end{center}

\centerline {Kh. P. Gnatenko \footnote{E-Mail address: khrystyna.gnatenko@gmail.com}}
\medskip
\centerline {\small \it Ivan Franko National University of Lviv, Department for Theoretical Physics,}
\centerline {\small \it 12 Drahomanov St., Lviv, 79005, Ukraine}
\centerline {\small \it Laboratory for Statistical Physics of Complex Systems,}
\centerline {\small \it Institute for Condensed Matter Physics, NAS of Ukraine 79011, Lviv, Ukraine}

\abstract{We study influence of noncommutativity of coordinates and noncommutativity of momenta on the motion of a particle (macroscopic body) in  uniform and non-uniform gravitational fields in noncommutative phase space of canonical type with preserved rotational symmetry.  It is shown that because of noncommutativity the motion of a particle  in gravitational filed is determined by its mass. The trajectory of motion of a particle in uniform gravitational field corresponds to the trajectory of harmonic oscillator with frequency determined by the value of parameter of momentum noncommutativity and mass of the particle. The equations of motion of a macroscopic body in gravitational filed depend on its mass and composition. From this follows violation of the weak equivalence principle caused by noncommutativity. We conclude that the weak equivalence principle is recovered in rotationally invariant noncommutative phase space if we consider the tensors of noncommutativity to be dependent on mass. So, finally we construct noncommutative algebra which is rotationally invariant, equivalent to noncommutative algebra of canonical type, and does not lead to violation of the weak equivalence principle.

}

\section{Introduction}

Much attention has been devoted recently to studies of features  of space structure at the Planck scale. To describe  space quantization the idea of noncommutativity of coordinates was considered.

In noncommutative phase space the following relations are satisfied
  \begin{eqnarray}
[X_{i},X_{j}]=i\hbar\theta_{ij},\label{form101}\\{}
[X_{i},P_{j}]=i\hbar(\delta_{ij}+\sum_k \frac{\theta_{ik}\eta_{jk}}{4}),\label{form1001}\\{}
[P_{i},P_{j}]=i\hbar\eta_{ij}.\label{form10001}{}
\end{eqnarray}
In the canonical version of noncommutative space  $\theta_{ij}$, $\eta_{ij}$ are elements of constant matrixes.

 Different problems were examined in noncommutative space among them free particles \cite{Djemai,BastosPhysA,Shyiko,Laba}, classical systems with various potentials \cite{Gamboa,Romero,Mirza,Djemai1,GnatenkoPLA14},  Landau problem  \cite{Gamboa1,Horvathy,Dayi,Horvathy05,Alvarez09,Daszkiewicz1}, many-particle systems \cite{Ho,Djemai,Daszkiewicz,GnatenkoPLA13,GnatenkoJPS13,Daszkiewicz2}, gravitational quantum well \cite{Bertolami1,Bastos}, and many others.

 Noncommutative algebra (\ref{form101})-(\ref{form10001}) with $\theta_{ij}$, $\eta_{ij}$ being elements of constant matrixes is not rotationally invariant. In noncommutative space of canonical type there is a problem of rotational symmetry breaking. Therefore different types of noncommutative algebras were considered to preserve the rotational symmetry (see, for instance, \cite{Moreno,Galikova,Amorim,GnatenkoPLA14}). Among them rotationally invariant noncommutative algebras with position-dependent noncommutativity  (see, for instance, \cite{Lukierski,Lukierski2009,BorowiecEPL,Borowiec,Borowiec1,Kupriyanov2009,Kupriyanov} and reference therein), with  involving spin degrees of freedom (see, for instance, \cite{Falomir09,Ferrari13,Deriglazov,Tkachuk}  and reference therein) have been widely studied.

  To preserve the rotational symmetry and construct rotationally invariant noncommutative algebra equivalent to noncommutative algebra of canonical type the generalization of parameters of noncommutativity to the tensors was considered \cite{GnatenkoIJMPA17}. The tensors were supposed to be constructed with the help of additional coordinates and momenta
  \begin{eqnarray}
\theta_{ij}=\frac{c_{\theta} l^2_{P}}{\hbar}\sum_k\varepsilon_{ijk}\tilde{a}_{k}, \ \
\eta_{ij}=\frac{c_{\eta}\hbar}{l^2_{P}}\sum_k\varepsilon_{ijk}\tilde{p}^b_{k}.\label{for130}
\end{eqnarray}
Here $l_P$ is the Planck length, $c_{\theta}$, $c_{\eta}$  are dimensionless constants,  $\tilde{a}_i$, $\tilde{b}_i$  $\tilde{p}^a_i$, $\tilde{p}^b_i$ are additional dimensionless coordinates and momenta conjugate to them which are governed by harmonic oscillators
 $ H^a_{osc}=\hbar\omega_{osc}\left({(\tilde{p}^{a})^{2}}+{\tilde{a}^{2}}\right)/2,$ $H^b_{osc}=\hbar\omega_{osc}\left({(\tilde{p}^{b})^{2}}+{\tilde{b}^{2}}\right)/2,$
with $\sqrt{{\hbar}/{m_{osc}\omega_{osc}}}=l_{P}$ and very large frequency $\omega_{osc}$ \cite{GnatenkoIJMPA17}.

Additional coordinates and additional momenta satisfy $[\tilde{a}_{i},\tilde{a}_{j}]=[\tilde{b}_{i},\tilde{b}_{j}]=[\tilde{a}_{i},\tilde{b}_{j}]=[\tilde{p}^{a}_{i},\tilde{p}^{a}_{j}]=[\tilde{p}^{b}_{i},\tilde{p}^{b}_{j}]=[\tilde{p}^{a}_{i},\tilde{p}^{b}_{j}]=0,$ $[\tilde{a}_{i},\tilde{p}^{a}_{j}]=[\tilde{b}_{i},\tilde{p}^{b}_{j}]=i\delta_{ij}$,
$[\tilde{a}_{i},\tilde{p}^{b}_{j}]=[\tilde{b}_{i},\tilde{p}^{a}_{j}]=0$. Also, $[\tilde{a}_{i},X_{j}]=[\tilde{a}_{i},P_{j}]=[\tilde{p}^{b}_{i},X_{j}]=[\tilde{p}^{b}_{i},P_{j}]=0$.
Therefore, $[\theta_{ij}, X_k]=[\theta_{ij}, P_k]=[\eta_{ij}, X_k]=[\eta_{ij}, P_k]=0$ as in the case of canonical noncommutativity ($\theta_{ij}$, $\eta_{ij}$ being constants). In this sense rotationally invariant algebra
\begin{eqnarray}
[X_{i},X_{j}]=ic_{\theta} l^2_{P} \sum_k\varepsilon_{ijk}\tilde{a}_{k},\label{rotinv}\\{}
[X_{i},P_{j}]=i\hbar\left(\delta_{ij}+\frac{c_{\theta}c_{\eta}}{4}({\bf \tilde{a}}\cdot{\bf \tilde{p}^{b}})\delta_{ij}-\frac{c_{\theta}c_{\eta}}{4}{\tilde a}_j{\tilde p}^{b}_i\right),\\{}
[P_{i},P_{j}]=\frac{c_{\eta}\hbar^2}{l_P^2}\sum_k\varepsilon_{ijk}\tilde{p}^{b}_{k}.{}\label{rotinv1}
\end{eqnarray}
is equivalent to noncommutative algebra of canonical type (\ref{form101})-(\ref{form10001})  \cite{GnatenkoIJMPA17}.

Influence of noncommutativity on the implementation of the equivalence principle was studied in the case of coordinates noncommutativity \cite{GnatenkoPLA13,GnatenkoJPS13,Saha,GnatenkoMPLA16,Saha1,Kryn18},  noncommutativity of coordinates and noncommutativity of momenta \cite{Bastos1,Bertolami2,GnatenkoPLA17}.
 The authors of paper \cite{Bertolami2} concluded that the equivalence principle holds  in the sense that an accelerated frame of reference is locally equivalent to
a gravitational field, unless noncommutative parameters are anisotropic ($\eta_{xy}\neq\eta_{xz}$). In our previous papers we studied the possibility to recover the weak equivalence principle in a space with noncommutativity of coordinates \cite{GnatenkoPLA13,GnatenkoMPLA16,Kryn18}, in four-dimensional noncommutative phase space of canonical type \cite{GnatenkoPLA17}.  In the present paper we study the implementation of the weak equivalence principle in rotationally-invariant noncommutative phase space (\ref{rotinv})-(\ref{rotinv1}). We examine features of motion of a particle (macroscopic body) in gravitational field in the space (\ref{rotinv})-(\ref{rotinv1}). The cases of uniform and non-uniform gravitational fields are studied and the dependence of the motion of a particle (macroscopic body) on its mass and composition is analyzed. We show that assumption that particles with different masses feel effect of noncommutativity with different parameters gives a possibility to preserve the weak equivalence principle.

The paper is organized as follows. In Section 2  features of description of composite system in rotationally invariant noncommutative phase space are presented. The motion of a particle (macroscopic body) in uniform gravitational filed in the noncommutative phase space is analyzed and the weak equivalence principle is considered in Section 3. In Section 4 the results are generalized to the case of non-uniform gravitational field. Conclusions are presented in Section 5.

\section{Description of many-particle system motion in noncommutative phase space with rotational symmetry}\label{rozd4}

Features of description of motion of composite system in rotationally invariant noncommutative phase space were studied in our previous paper {\cite{GnatenkoIJMPA18}}. In the paper we considered the general case when coordinates and momenta of different particles may satisfy noncommutative algebra with different tensors of noncommutativity $\theta^{(n)}_{ij}$ and $\eta^{(n)}_{ij}$
 \begin{eqnarray}
[X^{(n)}_{i},X^{(m)}_{j}]=i\hbar\delta_{mn}\theta^{(n)}_{ij},\label{ffor101}\\{}
[X^{(n)}_{i},P^{(m)}_{j}]=i\hbar\delta_{mn}\left(\delta_{ij}+\sum_k\frac{\theta^{(n)}_{ik}\eta^{(m)}_{jk}}{4}\right),\label{for1001}\\{}
[P^{(n)}_{i},P^{(m)}_{j}]=i\hbar\delta_{mn}\eta^{(n)}_{ij},\label{ffor10001}\\{}
\theta^{(n)}_{ij}=\frac{c_{\theta}^{(n)}l_P^2}{\hbar}\sum_k\varepsilon_{ijk}\tilde{a}_{k}, \ \
\eta^{(n)}_{ij}=\frac{c_{\eta}^{(n)}\hbar}{l_P^2}\sum_k\varepsilon_{ijk}\tilde{p}^b_{k}, \label{r13011}
 \end{eqnarray}
here indexes $m,n=(1...N)$ label the particles.

Coordinates and momenta which satisfy (\ref{ffor101})-(\ref{ffor10001}) can be represented as
\begin{eqnarray}
X^{(n)}_{i}=x^{(n)}_{i}-\frac{1}{2}\theta^{(n)}_{ij}p^{(n)}_j,\ \
P^{(n)}_{i}=p^{(n)}_{i}+\frac{1}{2}\eta^{(n)}_{ij}x^{(n)}_j.\label{rep2}{}
 \end{eqnarray}

The momenta and coordinates of the center-of-mass and  the momenta and the coordinates of the relative motion defined in the traditional way read ${\bf P}^c=\sum_{n}{\bf P}^{(n)}$, ${\bf X}^c=\sum_{n}\mu_{n}{\bf X}^{(n)}$, ${\bf\Delta P}^{(n)}={\bf P}^{(n)}-\mu_{n}{\bf P}^c,$ ${\bf \Delta X}^{(n)}={\bf X}^{(n)}-{\bf X}^c$,  here $\mu_n=m_n/M$, $M=\sum_{n=1}^{N}m_n$, coordinates ${\bf X}^{(n)}=(X^{(n)}_1,X^{(n)}_2,X^{(n)}_3)$ and momenta ${\bf P}^{(n)}=(P^{(n)}_1,P^{(n)}_2,P^{(n)}_3)$ satisfy (\ref{ffor101})-(\ref{ffor10001}). So, taking into account (\ref{ffor101})-(\ref{ffor10001}), we have the following commutation relations
\begin{eqnarray}
[X^c_i,X^c_j]=i\hbar\sum_{n}\mu_{n}^{2}\theta^{(n)}_{ij},\ \
[P^c_i,P^c_j]=i\hbar\sum_{n}\eta^{(n)}_{ij},\label{07}\\{}
[{X}^c_i,{P}^c_j]=i\hbar(\delta_{ij}+\sum_n\sum_k\mu_n\frac{\theta^{(n)}_{ik}\eta^{(n)}_{jk}}{4}).\label{010}\\{}
[{X}^c_{i},\Delta X_{j}^{(n)}]=i\hbar(\mu_{n}\theta^{(n)}_{ij}-\sum_m\mu_m^2{\theta}_{ij}^{(m)}),\label{0000}\\{} [{P}^c_i,\Delta{P}^{n}_j]=i\hbar(\eta^{(n)}_{ij}-\mu_n\sum_{m}\eta^{(m)}_{ij}).\label{000}
\end{eqnarray}

Because of relations (\ref{0000}), (\ref{000}) the motion of the center-of-mass is not independent of the relative motion in noncommutative phase space. In \cite{GnatenkoIJMPA18} we showed that in the case when tensors of noncommutativity corresponding to a particle are determined by its mass as
\begin{eqnarray}
\theta^{(n)}_{ij}=\frac{\tilde{\gamma} l^2_{P}}{\hbar m_n}\sum_k\varepsilon_{ijk}\tilde{a}_{k}, \ \
\eta^{(n)}_{ij}=\frac{\tilde{\alpha}\hbar m_n}{l^2_{P}}\sum_k\varepsilon_{ijk}\tilde{p}^b_{k},\label{efor130}
\end{eqnarray}
 namely when the following conditions hold
\begin{eqnarray}
c^{(n)}_{\theta}m_n=\tilde{\gamma}=const,\label{condt}\\
\frac{c^{(n)}_{\eta}}{m_n}=\tilde{\alpha}=const\label{conde}
\end{eqnarray}
where $\tilde{\gamma}$, $\tilde{\alpha}$ are constants which are the same for particles with different masses, one has
 \begin{eqnarray}\label{0}
 [{X}^c_{i},\Delta X_{j}^{(a)}]=[{P}^c_i,\Delta{P}^{a}_j]=0.
\end{eqnarray}

 Also, if conditions (\ref{condt}), (\ref{conde}) hold the commutation relations for coordinates and momenta of the center-of-mass read
\begin{eqnarray}
[X^c_i,X^c_j]=i\hbar{\theta}^c_{ij},\label{007}\ \
[P^c_i,P^c_j]=i\hbar{\eta}^c_{ij},\label{007}\\{}
[{X}^c_i,{P}^c_j]=i\hbar(\delta_{ij}+\sum_k\frac{\theta^c_{ik}\eta^c_{jk}}{4}).\label{0010}{}
\end{eqnarray}
 So, coordinates and momenta of the center-of-mass satisfy noncommutative algebra (\ref{form101})-(\ref{form10001}) with effective tensors of noncommutativity  which depend on the total mass of the system and do not depend on its composition.
\begin{eqnarray}
{\theta}^c_{ij}=\sum_n \mu^2_n\theta_{ij}^{(n)}=\frac{\tilde{\gamma} l^2_{P}}{\hbar M}\sum_k\varepsilon_{ijk}\tilde{a}_{k},\label{effc}\\
{\eta}^c_{ij}=\sum_{n}\eta^{(n)}_{ij}=\frac{\tilde{\alpha}\hbar M}{l^2_{P}}\sum_k\varepsilon_{ijk}\tilde{p}^b_{k}.\label{eff2c}
\end{eqnarray}
The noncommutative coordinates $X^c_i$ and noncommutative momenta ${P}^c_i$ can be represented as
\begin{eqnarray}
X^c_i=\sum_n\mu_n(x_i^{(n)}-\frac{1}{2}\theta^{(n)}_{ij}p^{(n)}_j)=x^c_i-\frac{1}{2}{\theta}^c_{ij}p^c_j,\label{xc}\\
P^c_i=\sum_n(p_i^{(n)}+\frac{1}{2}\eta^{(n)}_{ij}x^{(n)}_j)=p^c_i+\frac{1}{2}{\eta}^c_{ij}x^c_j,\label{pc}
\end{eqnarray}
here $x^c_i=\sum_n\mu_nx^{(n)}_i$, $p^c_i=\sum_n p^{(n)}_i$, coordinates and momenta $x^c_i$, $p^c_i$ satisfy the ordinary commutation relations.
Also, on the basis of the definition of $\Delta{\bf X}^{(n)}$, $\Delta{\bf P}^{(n)}$ using  (\ref{for130}), (\ref{rep2}) we obtain
\begin{eqnarray}
{\Delta X}^{(n)}_i=\Delta p_i^{(n)}+\frac{1}{2}\eta^{(n)}_{ij}\Delta x^{(n)}_j,\label{r1}\\
{\Delta P}^{(n)}_i=\Delta p_i^{(n)}+\frac{1}{2}\eta^{(n)}_{ij}\Delta x^{(n)}_j\label{r2}
\end{eqnarray}
where $\Delta x_i^{(n)}=x_i^{(n)}-x_i^c$, $\Delta p_i^{(n)}=p_i^{(n)}-\mu_n p_i^c$.

In the next sections these results will be used for studies of motion of macroscopic body in gravitational field in rotationally-invariant noncommutative phase space.

\section{Influence of noncommutativity on the motion in uniform gravitational filed}\label{rozd3}

First, let us study the motion of a particle of mass $m$ in noncommutative phase space with preserved rotational symmetry (\ref{rotinv})-(\ref{rotinv1}). For the particle in uniform gravitational filed one has the following Hamiltonian
\begin{eqnarray}
 H_p=\frac{{\bf P}^{2}}{2m}+mg X_1
\end{eqnarray}
here for convenience we choose the $X_1$ axis to be directed along the field direction. The coordinates and momenta $X_i$, $P_i$ satisfy (\ref{rotinv})-(\ref{rotinv1}).
 Because of dependence of tensors of noncommutativity on additional coordinates and momenta $\tilde{a}_i$, $\tilde{b}_i$  $\tilde{p}^a_i$, $\tilde{p}^b_i$ one has to consider the total Hamiltonian as follows
\begin{eqnarray}
 H=H_p+H^a_{osc}+H^b_{osc},\label{total}
\end{eqnarray}
where $H_{osc}^a$, $H_{osc}^b$ are Hamiltonians of harmonic oscillators. Using representation  (\ref{rep2}) one can write
\begin{eqnarray}
 H=\frac{{\bf p}^{2}}{2m}+mg x_1-\frac{({\bm \eta}\cdot{\bf L})}{2m}+ \frac{mg}{2}[{\bm \theta}\times {\bf p}]_1+\nonumber\\+\frac{[{\bm \eta}\times{\bf x}]^2}{8m}+H^a_{osc}+H^b_{osc}.\label{total}
\end{eqnarray}
Here we introduce the following notations ${\bm \theta}=(\theta_1,\theta_2,\theta_3)$,  ${\bm \eta}=(\eta_1,\eta_2,\eta_3)$,  $\theta_i=\sum_{jk}\varepsilon_{ijk}\theta_{jk}/2$, $\eta_i=\sum_{jk}\varepsilon_{ijk}\eta_{jk}/2$, and ${\bf L}=[{\bf x}\times{\bf p}]$.

It is convenient to rewrite the Hamiltonian  (\ref{total}) in the following form
$H=H_0+\Delta H$ with $H_0$, $\Delta H$ being defined as $H_0=\langle H_p\rangle_{ab}+H^a_{osc}+H^b_{osc}$,
$\Delta H= H-H_0=H_p-\langle H_p\rangle_{ab}.$  Notation  $\langle...\rangle_{ab}=\langle\psi^{a}_{0,0,0}\psi^{b}_{0,0,0}|...|\psi^{a}_{0,0,0}\psi^{b}_{0,0,0}\rangle$ is used for averaging over the degrees of freedom of harmonic oscillators $H_{osc}^a$, $H_{osc}^b$  in the ground states. The oscillators are considered to be in the ground states because the frequency $\omega_{osc}$ is supposed to be very large which leads to the large distance between their energy levels. So, harmonic oscillators in the ground states remain in them. Functions $\psi^{a}_{0,0,0}$, $\psi^{b}_{0,0,0}$ are well known eigenstates of harmonic oscillators $H^a_{osc}$, $H^b_{osc}$.
We have
\begin{eqnarray}
H_0=\frac{{\bf p}^{2}}{2m}+mg x_1+\frac{\langle\eta^2\rangle {\bf x}^2}{12m}+H^a_{osc}+H^b_{osc},\label{h0}\\
 \Delta H=-\frac{({\bm \eta}\cdot{\bf L})}{2m}+\frac{mg}{2}[{\bm \theta}\times {\bf p}]_1+\frac{[{\bm \eta}\times{\bf x}]^2}{8m}-\frac{\langle\eta^2\rangle {\bf x}^2}{12m},\label{delta}
\end{eqnarray}
where we take into account that $\langle\psi^{a}_{0,0,0}|\theta_i|\psi^{a}_{0,0,0}\rangle=0$, $\langle\psi^{b}_{0,0,0}|\eta_i|\psi^{b}_{0,0,0}\rangle=0$,
\begin{eqnarray}
\langle\theta_i\theta_j\rangle=\frac{c_{\theta}^2l_P^4}{\hbar^2}\langle\psi^{a}_{0,0,0}| \tilde{a}_i\tilde{a}_j|\psi^{a}_{0,0,0}\rangle=\frac{c_{\theta}^2l_P^4}{2\hbar^2}\delta_{ij}.\label{thetar2}\\
\langle\eta_i\eta_j\rangle= \frac{\hbar^2 c_{\eta}^2}{l_P^4}\langle\psi^{b}_{0,0,0}| \tilde{p}^{b}_i\tilde{p}^{b}_j|\psi^{b}_{0,0,0}\rangle=\frac{\hbar^2 c_{\eta}^2}{2 l_P^4}\delta_{ij}.\label{etar2}
\end{eqnarray}
and $\langle\theta_i\theta_j\rangle=\langle\theta^2\rangle\delta_{ij}/3$, $\langle\eta_i\eta_j\rangle=\langle\eta^2\rangle\delta_{ij}/3$.
In our paper \cite{GnatenkoIJMPA18} we showed that up to the second order in $\Delta H$ one can consider the hamiltonian $H_0$ because the corrections to the spectrum of $H_0$ up to the second order in the perturbation theory vanish.
Namely, in the first order of perturbation theory the corrections to the spectrum of $H_0$ caused by  $\Delta H$ read
\begin{eqnarray}
\Delta E^{(1)}=\langle\psi^p_{\{n_p\}}\psi^a_{0,0,0}\psi^b_{0,0,0}|\Delta H|\psi^p_{\{n_p\}}\psi^a_{0,0,0}\psi^b_{0,0,0}\rangle=\nonumber\\=\langle\psi^p_{\{n_p\}}|\langle H_p\rangle_{ab}-\langle H_p\rangle_{ab}|\psi^p_{\{n_p\}}\rangle=0.
\end{eqnarray}
Here $\psi^p_{\{n_p\}}$ are eigenfunctions of $\langle H_p\rangle_{ab}$ which corresponds to well known eigenfunctions of harmonic oscillator of mass $m$ and frequency $\sqrt{\eta}/\sqrt{6m}$ in uniform field $mg$ ($\{n_p\}$ being quantum numbers). In the second order of the perturbation theory we can write
\begin{eqnarray}
\Delta
E^{(2)}=\lim_{\omega_{osc}\rightarrow\infty}\nonumber\\\sum_{\{n_p^{\prime}\},\{n^{a}\},\{n^{b}\}}{\left|\left\langle \psi^{(0)}_{\{n_p^{\prime}\},\{n^{a}\},\{n^{b}\}}\left|
\Delta H\right|\psi^{(0)}_{\{n_p\},\{0\},\{0\}}\right\rangle\right|^{2}}\times\nonumber\\\times\left({E^p_{\{n_p^{\prime}\}}-E^p_{\{n_p\}}-\sum^{3}_{i=1}\hbar\omega_{osc}(n^{a}_{i}+n^{b}_{i})}\right)^{-1}\nonumber\\=0.\label{form311}
\end{eqnarray}
The sets of numbers $\{n_p^{\prime}\}$, $\{n^{a}\}$, $\{n^{b}\}$
 and $\{n_p\}$,$\{0\}$, $\{0\}$ do not coincide. Notation $E^p_{\{n_p\}}$ corresponds to the energy levels of $\langle H_p\rangle_{ab}$. In (\ref{form311}) we take into account that $\left|\left\langle \psi^{(0)}_{\{n_p^{\prime}\},\{n^{a}\},\{n^{b}\}}\left|
\Delta H\right|\psi^{(0)}_{\{n_p\},\{0\},\{0\}}\right\rangle\right|^{2}$ does not depend on $\omega_{osc}$ because  $\sqrt{{\hbar}/{m_{osc}\omega_{osc}}}=l_{P}$.

So, taking into account expression for $\Delta H$ (\ref{delta}), we can state that up to the second order in the parameters of noncommutativity for a particle in uniform gravitational filed one can consider Hamiltonian (\ref{h0}).
In this approximation the equations of motion of the particle read
\begin{eqnarray}
\dot x_i=\frac{p_i}{m},\ \
\dot p_i=-mg\delta_{i,1}-\frac{\langle\eta^2\rangle x_i}{6m}.\label{eqm1}
\end{eqnarray}
From  (\ref{eqm1}) the trajectory of a particle in uniform filed in rotationally invariant noncommutative phase space is the following
\begin{eqnarray}\label{tr}
x_i(t)=\left(x_{0i}+6g{\frac{m^2}{\langle\eta^2\rangle}}\delta_{1,i}\right)\cos\left(\sqrt{\frac{\langle\eta^2\rangle}{6m^2}}t\right)+\nonumber\\+\upsilon_{0i}\sqrt{\frac{6m^2}{\langle\eta^2\rangle}}\sin\left(\sqrt{\frac{\langle\eta^2\rangle}{6m^2}}t\right)-6g{\frac{m^2}{\langle\eta^2\rangle}}\delta_{1,i},
\end{eqnarray}
here $x_{0i}$, $\upsilon_{0i}$ are initial coordinates and velocities of the particle.
Note that the motion of the particle is affected only by the momentum noncommutativity. In the limit $\langle\eta^2\rangle\rightarrow0$ one obtains $x_i(t)=\delta_{1,i}gt^2/2+x_{0i}$ as it should be.
From (\ref{tr}) we have that particles with different masses move on different trajectories in uniform gravitational filed in rotationally invariant noncommutative phase space. From this we can conclude that the weak equivalence principle also known as the uniqueness of free fall principle is violated. The principle states that  velocity and position of a point mass in a gravitational field  are
independent of mass, composition and structure and depend only on its initial position and velocity.

We would like to stress that if condition (\ref{conde}) holds from (\ref{etar2}), (\ref{tr}), one has
\begin{eqnarray}\label{condit1}
\frac{\langle\eta^2\rangle}{m^2}=\frac{3\hbar^2\tilde{\alpha}^2}{2l_P^4}=B=const,\\
x_i(t)=\left(x_{0i}+{\frac{6g}{B}}\delta_{1,i}\right)\cos\left(\sqrt{\frac{B}{6}}t\right)+\nonumber\\+\upsilon_{0i}\sqrt{\frac{6}{B}}\sin\left(\sqrt{\frac{B}{6}}t\right)-{\frac{6g}{B}}\delta_{1,i}.\label{tr1}
\end{eqnarray}
constant $B$ does not depend on mass. So, if condition (\ref{conde}) is satisfied the trajectory of a particle in uniform gravitational filed does not depend on its mass as it has to be and the weak equivalence principle is recovered in noncommutative phase space with rotational symmetry.

The same conclusion can be done in the case of  motion of composite system  in the uniform gravitation field. For composite system (macroscopic body) of mass $M$ in uniform field we can write
\begin{eqnarray}
 H_s=\frac{( {\bf P}^c)^{2}}{2M}+MgX^{(c)}_1+H_{rel}\label{form777}
\end{eqnarray}
Here ${\bf P}^c$, ${\bf X^{(c)}}$ are momenta and coordinates of the center-of-mass of the composite system (macroscopic body). Hamiltonian $H_{rel}$ corresponds to the relative motion.
As was shown in the previous section if conditions (\ref{condt}), (\ref{conde}) hold the coordinates and the momenta of the center-of-mass and the coordinates and the momenta of the relative motion satisfy (\ref{0})-(\ref{0010}), and can be represented as (\ref{xc})-(\ref{r2}).
So, we can write
\begin{eqnarray}
H_0=\frac{({\bf p}^c)^{2}}{2M}+Mg x^c_1+\frac{\langle(\eta^c)^2\rangle ({\bf x}^c)^2}{12M}+\langle H_{rel}\rangle_{ab}+\nonumber\\+H^{(a)}_{osc}+H^{(b)}_{osc}.
\end{eqnarray}
Note  that $[H_0,\langle H_{rel}\rangle_{ab}]=0$, because  $\langle H_{rel}\rangle_{ab}$ depends on $\Delta x^{(n)}_i$, $\Delta p^{(n)}_i$ which commute with $x_c^i$ and $p^c_i$ and ${\tilde a}_i$, ${\tilde b}_i$,${\tilde p}_i^a$, ${\tilde p}_i^b$. So, the trajectory of the center-of-mass of composite system is as follows
\begin{eqnarray}\label{trcm}
x^c_i(t)=\left(x^c_{0i}+6g{\frac{M^2}{\langle(\eta^c)^2\rangle}}\delta_{1,i}\right)\cos\left(\sqrt{\frac{\langle(\eta^c)^2\rangle}{6M^2}}t\right)+\nonumber\\+\upsilon^c_{0i}\sqrt{\frac{6M^2}{\langle(\eta^c)^2\rangle}}\sin\left(\sqrt{\frac{\langle(\eta^c)^2\rangle}{6M^2}}t\right)-6g{\frac{M^2}{\langle(\eta^c)^2\rangle}}\delta_{1,i},
\end{eqnarray}
If relation (\ref{conde}) does not hold, we have that the trajectory of the center-of-mass of a body depends on its mass. When relation (\ref{conde}) is satisfied the effective tensor of momentum noncommutativity  is defined as (\ref{eff2c}) and we can write
\begin{eqnarray}
\frac{{\langle(\eta^c)^2\rangle}}{M^2}=\frac{3\hbar^2\tilde{\alpha}^2}{2l_P^4}=B=const,\\
x^c_i(t)=\left(x^c_{0i}+{\frac{6g}{B}}\delta_{1,i}\right)\cos\left(\sqrt{\frac{B}{6}}t\right)+\nonumber\\+\upsilon_{0i}\sqrt{\frac{6}{B}}\sin\left(\sqrt{\frac{B}{6}}t\right)-{\frac{6g}{B}}\delta_{1,i},
\end{eqnarray}
So, the trajectory of a body in uniform filed does not depend on its mass and composition because of the relation (\ref{conde}).

Note also that taking into account (\ref{tr}), the motion of the center-of-mass of a system of $N$ non-interacting particles with masses $m_a$ in uniform gravitational field is described by the following trajectory
\begin{eqnarray}\label{tra}
x^c_i(t)=\sum_a\mu_ax^{(a)}_i(t)=-\sum_a6g\mu_a{\frac{m_a^2}{\langle(\eta^{(a)})^2\rangle}}\delta_{1,i}+\nonumber\\+\sum_a\mu_a\left(x^{(a)}_{0i}+6g{\frac{m_a^2}{\langle(\eta^{(a)})^2\rangle}}\delta_{1,i}\right)\cos\left(\sqrt{\frac{\langle(\eta^{(a)})^2\rangle}{6m_a^2}}t\right)+\nonumber\\+\sum_a\mu_a\upsilon^{(a)}_{0i}\sqrt{\frac{6m_a^2}{\langle(\eta^{a})^2\rangle}}\sin\left(\sqrt{\frac{\langle(\eta^{a})^2\rangle}{6m_a^2}}t\right),
\end{eqnarray}
where index $a$ labels the particles, $x^{(a)}_{0i}$, $\upsilon^{(a)}_{0i}$ are initial coordinates and initial velocities of the particle with mass $m_a$. It is important to mention that if relation (\ref{conde}) holds, using $x^{(c)}_{0i}=\sum_a\mu_ax^{(a)}_{0i}$ and $\upsilon^{(c)}_{0i}=\sum_a\mu_a\upsilon^{(a)}_{0i}$, expression (\ref{tra}) reduces to (\ref{trcm}).

\section{Motion in the non-uniform gravitational field and the weak equivalence principle}
Let us study Hamiltonian which corresponds to a particle of mass $m$ in non-uniform gravitational field
\begin{eqnarray}
H_p=\frac{{P}^2}{2m}-\frac{G{\tilde M}m}{X},
 \end{eqnarray}
here $X=|{\bf X}|=\sqrt{\sum_i X_i^2}$. Coordinates $X_i$ and momenta $P_i$ satisfy relations (\ref{rotinv})-(\ref{rotinv1}). Using representation (\ref{rep2}) we can write
$X=\sqrt{x^2-({\bm \theta}\cdot{\bf L})+[{\bm \theta}\times{\bf p}]^2/4}$. Up to the second order in the parameters of noncommutativity the Hamiltonian $H_p$ reads
\begin{eqnarray}
H_p=\frac{{p}^2}{2m}-\frac{G{\tilde M}m}{x}-\frac{({\bm \eta}\cdot{\bf L})}{2m}+\frac{[{\bm \eta}\times{\bf x}]^2}{8m}-\nonumber\\-\frac{G{\tilde M}m}{2x^3}({\bm \theta}\cdot{\bf L})-\frac{3G{\tilde  M}m}{8x^5}({\bm \theta}\cdot{\bf L})^2+\frac{G{\tilde M}m}{16}\times\nonumber\\ \times \left(\frac{1}{x^2}[{\bm \theta}\times{\bf p}]^2\frac{1}{x}+\frac{1}{x}[{\bm \theta}\times{\bf p}]^2\frac{1}{x^2}+\frac{\hbar^2}{x^7}[{\bm \theta}\times{\bf x}]^2\right),\label{hs}
 \end{eqnarray}
$x=|{\bf x}|$. Note, that the last term in (\ref{hs}) appears because of noncommutativity of operators $x^2$ and $[{\bm \theta}\times{\bf p}]^2$ under square root in $X$. The details of calculation of expansion of $1/X$ over the parameters of noncommutativity can be found in  \cite{GnatenkoPLA14}.
So, one can write expression for $\Delta H$
 \begin{eqnarray}
\Delta H=-\frac{({\bm \eta}\cdot{\bf L})}{2m}+\frac{[{\bm \eta}\times{\bf x}]^2}{8m}-\frac{\langle\eta^2\rangle { x}^2}{12m}-\frac{G{\tilde  M} m}{2x^3}({\bm \theta}\cdot{\bf L})+\nonumber\\+\frac{G{\tilde  M}m L^2\langle\theta^2\rangle}{8 x^5}+\frac{G{\tilde  M} m}{16}\left(\frac{1}{x^2}[{\bm \theta}\times{\bf p}]^2\frac{1}{x}+\right.\nonumber\\\left.+\frac{1}{x}[{\bm \theta}\times{\bf p}]^2\frac{1}{x^2}+\frac{\hbar^2}{x^7}[{{\bm \theta}}\times{\bf x}]^2\right)-\frac{3G{\tilde M}m}{8x^5}({\bm \theta}\cdot{\bf L})^2-\nonumber\\-\frac{G{\tilde  M}m\langle\theta^2\rangle}{24}\left(\frac{1}{x^2} p^2\frac{1}{x}+\frac{1}{x}p^2\frac{1}{x^2}+\frac{\hbar^2}{x^5}\right).\label{dd}
 \end{eqnarray}
Similarly as in the case of uniform gravitational field up to the second order in parameters of noncommutativity one can study the following Hamiltonian
\begin{eqnarray}
H_0=\frac{{ p}^2}{2m}-\frac{G{\tilde  M}m}{x}+\frac{\langle\eta^2\rangle  x^2}{12m}-\frac{G{\tilde  M}m L^2\langle\theta^2\rangle}{8 x^5}+\nonumber\\+\frac{G{\tilde  M}m\langle\theta^2\rangle}{24}\left(\frac{2}{x^3} p^2+\frac{6i\hbar}{x^5}({\bf x}\cdot{\bf p}) -\frac{\hbar^2}{x^5}\right)+\nonumber\\+H^a_{osc}+H^b_{osc}.\label{he}
\end{eqnarray}
 From (\ref{he}), the equations of motion of a particle in non-uniform gravitational filed read
\begin{eqnarray}
\dot{{\bf x}}=\frac{{\bf p}}{m}-\frac{G{\tilde  M}m\langle\theta^2\rangle}{12}\left(\frac{1}{x^3}{\bf p}-\frac{3{\bf x}}{x^5}({\bf x}\cdot{\bf p})\right),\label{r}\\
\dot{{\bf p}}=-\frac{G{\tilde  M}m {\bf x}}{x^3}-\frac{\langle\eta^2\rangle  {\bf x}}{6m}-\frac{G{\tilde M}m\langle\theta^2\rangle}{4}\left(\frac{1}{x^5}({\bf x}\cdot{\bf p}){\bf p}-\right.\nonumber\\\left.-\frac{2{\bf x}}{x^5}p^2+\frac{5{\bf x}}{2x^7}L^2+\frac{5\hbar^2{\bf x}}{6x^7}-\frac{5i\hbar}{x^7}{\bf x}({\bf x}\cdot{\bf p})\right).\label{p}
\end{eqnarray}
In the limit $\hbar\rightarrow0$ from (\ref{r}), (\ref{p}) we obtain
\begin{eqnarray}
\dot{{\bf x}}={\bm \upsilon}-\frac{G{\tilde  M}m^2\langle\theta^2\rangle}{12}\left(\frac{1}{x^3}{\bm \upsilon}-\frac{3{\bf x}}{x^5}({\bf x}\cdot{\bm \upsilon}) \right),\label{rv}\\
\dot{{\bm \upsilon}}=-\frac{G {\tilde M} {\bf x}}{x^3}-\frac{\langle\eta^2\rangle {\bf x}}{6m^2}-\nonumber\\-\frac{G {\tilde M} m^2\langle\theta^2\rangle}{4}\left(\frac{1}{x^5}({\bf x}\cdot{\bm \upsilon}){\bm \upsilon}-\frac{2{\bf x}}{x^5}\upsilon^2+\frac{5{\bf x}}{2x^7}[{\bf x}\times{\bm\upsilon}]^2\right),\label{pv}
\end{eqnarray}
where the vector ${\bm \upsilon}={{\bf p}}/{m}$ is introduced.
The obtained equations of motion depend on the products $m^2\langle\theta^2\rangle$ and $\langle\eta^2\rangle/m^2$.
Supposing that conditions (\ref{condt}), (\ref{conde}) hold we have
\begin{eqnarray}
\dot{{\bf x}}={\bm \upsilon}-\frac{G{\tilde  M}A}{12}\left(\frac{1}{x^3}{\bm \upsilon}-\frac{3{\bf x}}{x^5}({\bf x}\cdot{\bm \upsilon}) \right),\label{ra}\\
\dot{{\bm \upsilon}}=-\frac{G {\tilde M} {\bf x}}{x^3}-\frac{B{\bf x}}{6}-\nonumber\\-\frac{G {\tilde M} A}{4}\left(\frac{1}{x^5}({\bf x}\cdot{\bm \upsilon}){\bm \upsilon}-\frac{2{\bf x}}{x^5}\upsilon^2+\frac{5{\bf x}}{2x^7}[{\bf x}\times{\bm\upsilon}]^2\right).\label{pa}
\end{eqnarray}
Here we use (\ref{condit1}) and take into account that on condition (\ref{condt}) the following relation is satisfied
\begin{eqnarray}\label{condit2}
{\langle\theta^2\rangle}{m^2}=\frac{3\alpha^2l_P^4m^2}{2\hbar^2}=A=const
\end{eqnarray}
where $A$  is a constant which does not depend on mass.
Obtained equations of motion (\ref{ra}), (\ref{pa}) depends on constants $A$, $B$ which are the same for different particles. So, the weak equivalence principle is recovered in noncommutative phase space with preserved rotational symmetry due to relations (\ref{condt}), (\ref{conde}).

In quantum case, if (\ref{condt}), (\ref{conde}) are satisfied  the equations  (\ref{r}), (\ref{p}) read
\begin{eqnarray}
\dot{{\bf x}}={\bm \upsilon}-\frac{G {\tilde  M} B}{12}\left(\frac{1}{x^3}{\bm \upsilon}-\frac{3{\bf x}}{x^5}({\bf x}\cdot{\bm \upsilon}) \right),\label{rvx}\\
\dot{{\bm \upsilon}}=-\frac{G {\tilde  M} {\bf x}}{x^3}-\frac{B{\bf x}}{6}-\frac{G {\tilde M} A}{4}\left(\frac{1}{x^5}({\bf x}\cdot{\bm \upsilon}){\bm \upsilon}-\frac{2{\bf x}}{x^5}\upsilon^2+\right.\nonumber\\\left.+\frac{5{\bf x}}{2x^7}[{\bf x}\times{\bm\upsilon}]^2+\frac{5\hbar^2{\bf x}}{6m^2x^7}-\frac{5i\hbar}{mx^7}{\bf x}({\bf x}\cdot{\bm \upsilon})\right).\label{pvx}
\end{eqnarray}
 The equations depend on ${\hbar}/{m}$, as it has to be. This is because of commutation relation $[{\bf x}, {\bm \upsilon}]=i\hbar\hat{I}/m$ \cite{Greenberger68}.

So, in the case of preserving of  conditions (\ref{condt}), (\ref{conde}), the motion of a particle in nonuniform gravitational field does not depend on its mass as it is in the ordinary space.

The conclusion can be generalized to the case of motion of a composite system (macroscopic body) in non-uniform gravitational filed in noncommutative phase space with rotational symmetry. Similarly as was shown in the previous section for motion of a composite system in non-uniform gravitational field up to the second order in the parameters of noncommutativity we can write the following Hamiltonian

\begin{eqnarray}
H_s=\frac{{(P^c)}^2}{2M}-\frac{G{\tilde M}M}{(X^c)^2}+H_{rel},\\
H_0=\frac{{ (p^c)}^2}{2M}-\frac{G{\tilde  M} M}{x^c}+\frac{\langle(\eta^c)^2\rangle  (x^c)^2}{12M}-\nonumber\\-\frac{G{\tilde M} M (L^c)^2\langle\theta^2\rangle}{8 (x^c)^5}+\frac{G{\tilde M}M\langle(\theta^c)^2\rangle}{24}\left(\frac{2}{(x^c)^3} (p^c)^2+\right.\nonumber\\\left.
+\frac{6i\hbar}{(x^c)^5}({\bf x}^c\cdot{\bf p}^c) -\frac{\hbar^2}{(x^c)^5}\right)+\langle H_{rel}\rangle_{ab}+H^a_{osc}+H^b_{osc}\nonumber\\
\end{eqnarray}
Here we use (\ref{xc})-(\ref{r2}).
 Taking into account (\ref{condt}), (\ref{conde}), the equations of motion do not depend on the composition of a system and on its mass and they read
\begin{eqnarray}
\dot{{\bf x}}^c={\bm \upsilon}^c-\frac{G{\tilde  M}B}{12}\left(\frac{1}{(x^c)^3}{\bm \upsilon}^c-\frac{3{\bf x}^c}{(x^c)^5}({\bf x^c}\cdot{\bm \upsilon}^c) \right),\label{rac}\\
\dot{{\bm \upsilon}^c}=-\frac{G {\tilde M} {\bf x}^c}{(x^c)^3}-\frac{B{\bf x}^c}{6}-\frac{G {\tilde M} A}{4}\left(\frac{1}{(x^c)^5}({\bf x}^c\cdot{\bm \upsilon}^c){\bm \upsilon}^c-\right.\nonumber\\\left.-\frac{2{\bf x}^c}{(x^c)^5}(\upsilon^c)^2+\frac{5{\bf x}^c}{2(x^c)^7}[{\bf x}^c\times{\bm\upsilon}^c]^2\right).\label{pac}
\end{eqnarray}
We would like to stress that if conditions (\ref{condt}), (\ref{conde}) are not satisfied the equations of motion of a body (composite system) depends on its mass and $\langle(\theta^c)^2\rangle$, $\langle(\eta^c)^2\rangle$ which according to definitions  ${\theta}^c_{ij}=\sum_n \mu^2_n\theta_{ij}^{(n)}$, ${\eta}^c_{ij}=\sum_{n}\eta^{(n)}_{ij}$ (see (\ref{effc}), (\ref{eff2c})) depend on the composition. This dependence is additional fact which causes violation of the weak equivalence principle.

 At the end of this section we would like to note that conditions  (\ref{condt}),  (\ref{conde}) are in agreement with the conditions which were proposed to solve the list of problems in four dimensional  (2D configurational space and 2D momentum space) noncommutative phase space of canonical type in \cite{GnatenkoPLA17,GnatenkoMPLA17}. Namely in the papers we proposed the parameter of coordinate noncommutativity to be proposed inversely to mass, and parameter of momentum noncommutativity to be proposed to mass. The same dependence we have in the case of rotationally-invariant noncommutative space for the tensors of noncommutativity  (\ref{efor130}).

\section{Conclusions}
In the paper we studied the influence of noncommutativity on the motion in gravitational filed in rotationally invariant space with noncommutativity of coordinates and noncommutativity of momenta (\ref{rotinv})-(\ref{rotinv1}). The rotationally invariant noncommutative algebra is constructed with the help of generalization of parameters of noncommutativity to tensors  (\ref{for130}).

Motion of a particle in uniform gravitational filed was studied. We showed that in rotationally invariant noncommutative phase space up to the second order in the parameters of noncommutativity this motion is affected only  by the noncommutativity of momenta. The trajectory of a particle in uniform filed corresponds to the trajectory of motion of harmonic oscillator and depends on the particle mass and the value of parameter of momentum noncommutativity (\ref{tr}).  It was concluded that in the case when particles with different masses feel noncommutativity with different parameters, namely when condition (\ref{conde}) is satisfied the trajectory of a particle does not depend on mass (\ref{tr1})   and the weak equivalence principle is recovered in rotationally invariant noncommutative phase space. The same conclusion was done for the case of motion of a particle in non-uniform gravitational filed. We showed that due to relations  (\ref{condt}), (\ref{conde}) quantum equations of motion of the particle depend on $\hbar/m$ (\ref{rvx}),(\ref{pvx}), as it has to be, and in the classical limit they do not depend on mass (\ref{ra}), (\ref{pa}).

 Besides the conditions  (\ref{condt}), (\ref{conde})  are important in consideration of motion of composite system (macroscopic body) in rotationally invariant noncommutative phase space. Namely, preserving of  relations (\ref{condt}), (\ref{conde}) leads to commutativity of coordinates of the center-of-mass and relative coordinates, momenta the center-of-mass and relative momenta (\ref{0}) and to recovering independence of motion of the center-of-mass of composite system in gravitational field on its mass and composition (\ref{rac}), (\ref{pac}). So, the noncommutative algebra (\ref{ffor101})-(\ref{ffor10001}) with (\ref{efor130}) is rotationally invariant and and does not lead to violation of the weak equivalence principle.

We would like also to mention here another important conclusion which can be done due to  relations (\ref{condt}), (\ref{conde}). In our paper \cite{GnatenkoIJMPA18} we showed that when conditions (\ref{condt}), (\ref{conde}) are satisfied, the noncommutative coordinates can be considered as kinematic variables and noncommutative momenta are proportional to mass; the noncommutative algebra for coordinates and momenta of the center-of-mass corresponds to noncommutative algebra for coordinates and momenta of individual particles with effective parameters of noncommutativity (\ref{007})-(\ref{0010}).

So, the idea of dependence of parameters of noncommutativity on the mass (\ref{condt}), (\ref{conde}) gives possibility to solve the list of problems in rotationally invariant noncommutative phase space among them violation of the weak equivalence principle. In addition the idea is important not only in noncommutative space. In  deformed space with minimal length $[X,P]=i\hbar(1+\beta P^2)$ special dependence  of the parameter of deformation $\beta$ on mass leads to recovering of the weak equivalence principle, preserving of the properties of the kinetic energy, independence of Galilean and Lorentz transformations of mass  \cite{Tk1,Tk2,Tk3}.

\section*{Acknowledgments}
The author thanks Prof. V. M. Tkachuk and Dr. Yu. S. Krynytskyi for their
advices and great support during research studies. This work was partly supported by the grant of the President of Ukraine for support of scientific researches of young scientists (F-75).

\end{document}